# Theoretical insights into charge transfer plasmon lifetime

Alemayehu Nana Koya*, Longnan Li and Wei Li*

*GPL Photonics Laboratory, State Key Laboratory of Luminescence and Applications, Changchun Institute of Optics, Fine Mechanics and Physics, Chinese Academy of Sciences, Changchun 130033, Jilin, P. R. China*

E-mail: alemayehu.koya@gmail.com; weili1@ciomp.ac.cn

**Abstract**

Understanding the spectral and temporal dynamics of charge transfer plasmon resonances that emerge in conductively connected plasmonic nanoparticles is crucial for exploiting their potentials for enhanced infrared spectroscopy and optical computing. In this article, we present a theoretical study based on classical electromagnetism to describe the spectral signature and dephasing time of charge transfer plasmons. By fitting the scattering curves and near-field amplitude oscillations, we determine the spectral linewidth and lifetime of charge transfer plasmons in conductively connected gold nanodisk dimers. We find that, compared with the well-known particle plasmons and dimer plasmons, charge transfer plasmons have a longer lifetime, which can be further extended by manipulating the geometric parameters of nanojunction and nanoparticles. Moreover, quantitative analyses of the optical near-field amplitude reveal that charge transfer plasmon modes oscillate completely out of phase with particle plasmon and dimer plasmon modes. The dephasing time and charge transfer rate are found to be on a few femtosecond timescale, implying that conductively connected plasmonic nanoparticles hold great promise as channels for coherent transfer of energy and information in future all-optical computing devices.

1. Introduction

As a result of interaction of the elementary excitations of constituent nanoparticles, strongly coupled plasmonic nanosystems exhibit non-trivial optical responses [1, 2]. These include multimodal resonances, asymmetric lineshapes, and complex temporal dynamics [3 - 5]. In particular, charge transfer plasmon (CTP) modes that emerge in conductively coupled particles (CCPs) have interesting spectral features including ultra-narrow line-widths and broadly tunable infrared resonances [6]. The CTP mode in CCP appears as a result of electron transport across conductive junction [7] or through quantum tunneling [8]. The resonance peak of CTP mode can be sensitivity tuned by modulating the junction conductance [9, 10]. Moreover, in addition to the near-infrared CTP resonance, CCPs can also sustain additional plasmonic mode called screened bonding dimer plasmon (sBDP) resonance in the visible range [11]. As a result, conductively coupled plasmonic nanostructures have found applications in molecular sensing [12], refractometry [13, 14], enhanced spectroscopy [15], and optoelectronics [16].

Some of plasmon enhanced applications, like sensing, benefit from prolonged plasmon lifetime so that reduction of plasmon damping has been subject of extensive research for a couple of decades [17, 18]. In principle, the processes that lead to plasmon damping include radiative decay via transformation of plasmons into photons, non-radiative decay occurring through excitation of electron-hole pairs, and surface decay which is associated with electron scattering off the particle surface and interaction with environment. The speed of plasmon damping is characterized by the time constant $T_2$, which is related to the plasmon resonance linewidth $\Gamma$ according to $T_2 = 2\hbar/\Gamma$, where $\hbar$ is Planck's reduced constant. However, plasmon decay occurs on the order of 5−20 fs, making it difficult to directly measure via time-resolved methods [19, 20].

In this regard, the dephasing time of particle plasmons of single metallic nanostructures has been extensively explored both theoretically and experimentally [17, 21, 22]. Recent years have seen significant research efforts to control and enhance dephasing time of gap plasmons in coupled nanostructures of various configurations [23 - 25]. Nevertheless, the temporal dynamics of CTP modes of conductively coupled plasmonic nanoparticles remains unexplored and the underlying physics behind the unique optical responses of CCPs remains elusive. This knowledge gap underscores the necessity of conductively coupled plasmonic nanoparticles as building blocks of optoelectronic devices and sensors.

In this article, we theoretically examine the spectral features and temporal dynamics of charge transfer plasmons in conductively coupled gold nanodisk dimers. In particular, we conduct a detailed study of the dephasing time of CTP mode compared with localized surface plasmon (LSP) resonance of equivalent nanorod and bonding dimer plasmon (BDP) mode of unlinked nanodimer. We find that, CTP mode has narrower linewidth and longer dephasing time. We show that the dephasing time of charge transfer plasmon resonances can be further manipulated by controlling the conductance of nanojunctions and size of nanoparticles. Moreover, the optical near-field oscillation of connected plasmonic nanoparticles reveal that CTP modes oscillate completely out of phase with both LSP and BDP modes. These findings have implications for utilizing interconnected plasmonic nanostructures as building blocks of molecular sensors and optoelectronic devices.

2. **Theory and Methods**

One of the key factor that determines charge transfer plasmon resonance in linked nanoparticles is the conductance of the linker, which is determined by [26]

$$G = \sigma \frac{A}{l} \quad (1)$$

where, σ is conductivity, *A* is cross section and *l* is length of the nanojunction. The threshold conductance required to sustain CTP mode in conductively bridged nanoparticle dimers is given by [11]

$$G_{CTP} = \frac{\omega_{CTP}}{16\pi} \frac{D^2}{l} \quad (2)$$

where, $\omega_{CTP}$ is the charge transfer plasmon frequency and $D$ is the nanoparticle diameter. For a small junction, one can assume the surface charge to be constant across the junction, resulting in a total charge Q to be transported in each cycle. The characteristic charge transfer time can be estimated as the time when one nanoparticle receives half of the initial energy from other nanoparticle and it happens when $\frac{\omega_{CTP}}{2}t = \frac{\pi}{4}$ [27]. This equation sets up the charge transfer time $\tau_{CT}$ [9]

$$t_{CT} = \frac{\pi}{2\omega_{CTP}} \qquad (3)$$

These equations imply that CTP resonance and charge transport speed can be easily manipulated by changing the geometric parameters of the nanolinker and the nanoparticle.

To explore the spectral linewidth and dephasing time of CTP mode, we used conductively connected nanodisk dimer linked by conductive junction (see **Figure 1(a)**). For the sake of comparison of the spectral and temporal signatures of CTP with the well-known LSP mode, initially, initially, we set mode volumes of both linked nanodimer and nanorod equivalent ($V \approx 5.3154 \times 10^5 nm^3$). This leads to set the geometric parameters of the studied nanostructures as follows. The conductively linked disk nanodimer is characterized by its junction parameters (length $l$ = 30 nm, width w = 22 nm) and nanodisk parameters (diameter D = 97 nm, and height H = 35 nm). And the high aspect ratio nanorod has a length of L = 320 nm and diameter d = 46 nm. However, at the later stage, to investigate the geometric effects of nanojunction and nanoparticles on the spectral and temporal signatures of CTP mode, we varied the nanojunction volume and nanoparticle diameter in the linked nanodisk dimers.

To calculate the optical responses of the studied nanosystems, we employed the finite-difference time-domain (FDTD) simulation method with smallest mesh sizes ($2 \times 2 \times 2\ nm^3$) to ensure highest simulation accuracy. The optical constants of gold were taken from the experimental data of Johnson and Christy [28]. The scattering intensities of the studied nanostructures were calculated using the total-field-scattered-field (TFSF) technique. The local field properties were simulated at resonant wavelengths and the electric near-field amplitudes were acquired by placing point monitor at mid-gap of the nanodimer and at 1 nm away from the nanorod and nanojunction of the linked nanodimer.

The plasmon resonance linewidths of the investigated nanosystems were determined by Lorentz fitting of their scattering spectra. To determine the dephasing time $T_2$ of the studied plasmonic modes, we used Heisenberg's time-energy relationship $T_2 = 2\hbar/\Gamma$. These indirectly determined dephasing times of the studied plasmonic modes acquired from scattering spectra were compared with the optical near-field lifetimes obtained by fitting the simulation data of the near-field amplitude decay using the exponential decay function of $E(t) = E_0 e^{-t/\tau}$, where $E_0$ is constant and $\tau$ is the lifetime of the near-field [24].

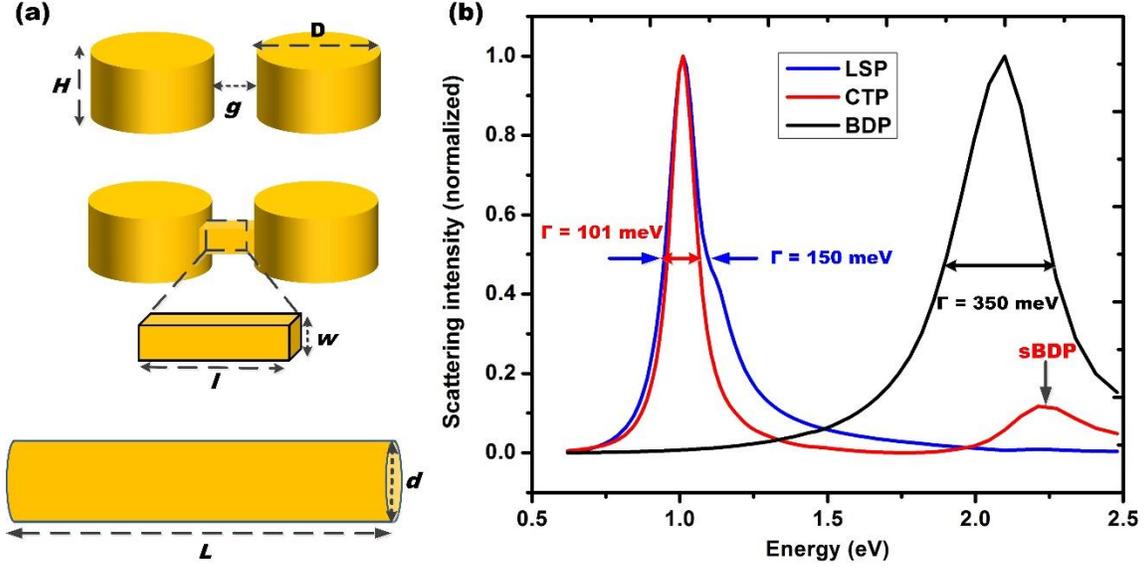

**Figure 1**. (a) Design of studied nanosystems: strongly coupled nanodisk dimer ($D$ = 97 nm, $H$ = 35 nm, g = 30 nm), conductively linked disk nanodimer ($l$ = 30 nm, $w$ = 22 nm), and high aspect ratio Au nanorod ($L$ = 320 nm and $d$ = 46 nm). (b) Spectral signatures of bonding dimer plasmon (BDP) mode of the disk nanodimer, charge transfer plasmon (CTP) mode of linked nanodimer and localized surface plasmon (LSP) mode of nanorod.

### 3. Results and Discussion

*Spectral and temporal signatures of CTP mode:* For the studied nanostructure geometries displayed in **Figure 1(a)**, both the linked nanodimer and nanorod have an equivalent mode volume so that the corresponding CTP and LSP modes have comparable resonance energies ($\hbar\omega_{res}$ = 1.01 and 1.02 eV respectively) (see **Figure 1(b)**). In addition to the CTP mode, the linked nanodimer also supports screened bonding dimer plasmon (sBDP) mode (around 2.21 eV) that comes from strong coupling of the nanodisks (see black arrow). Nevertheless, the spectral linewidth of CTP mode obtained by Lorentz fitting is found to be narrower ($\Gamma_{CTP} = 101.82\ meV$) than that of LSP resonance ($\Gamma_{LSP} = 149.84\ meV$). The spectral broadening observed in the LSP resonance is due to the excitation of dissipative dipole plasmon mode that can be excited in such high aspect ratio nanorods [6]. Similarly, the dephasing time obtained using $T_2 = 2\hbar/\Gamma$ implies that CTP mode lasts longer than that of LSP mode ($T_2^{CTP} = 12.92\ fs, T_2^{LSP} = 8.78\ fs$). On the other hand, even though 2.73 % reduction in the mode volume of nanodimer via removing the junction, BDP mode of unlinked nanodimer has broadest linewidth ($\Gamma_{BDP} = 0.353\ eV$), lowest quality factor ($Q = \hbar\omega_{res}/\Gamma = 5.88$), and shortest dephasing time ($T_2^{BDP} = 3.72\ fs$). For such small dimer gaps (g = 30 nm), strong near-field enhancement and short dephasing time are expected due to plasmon localization, which leads to enhanced radiative decay [23].

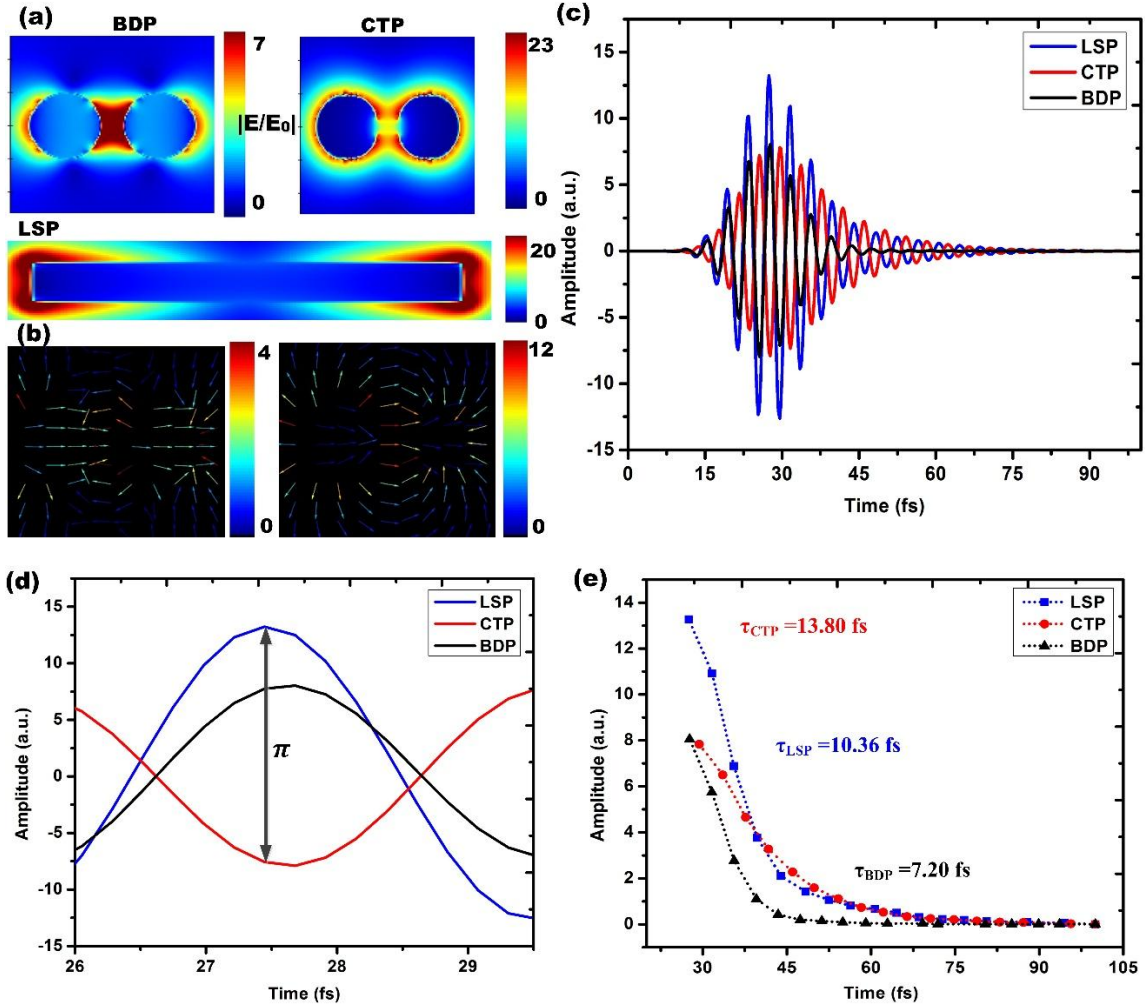

**Figure 2.** (a) Resonantly calculated near-field profiles of LSP, CTP and BDP modes. (b) Corresponding charge distribution maps of BDP and CTP modes. (c) Comparative electric near-field amplitudes of the three modes calculated at 1 nm for the nanorod and linked dimer and at the mid-gap of unlinked nanodimer. (d) Zoom in view of phase differences in the electric near-field oscillations of CTP, and LSP and BDP modes shown in (c). (e) Fitting of exponential decays of LSP, CTP and BDP resonances.

To have a better understanding of the distinct plasmonic modes that emerge in the studied nanostructures with comparable mode volumes, we have simulated the hot spot maps, charge distributions and electric near-field amplitudes using near-field monitors. In particular, resonantly calculated local field profiles and charge distribution maps show that bare nanodimer and nanorod have a dipole type distribution whereas the linked nanodimer exhibits steady charge flow across the nanojunction (**Figure 2(a, b)**). Moreover, the electric near-field amplitudes of the three modes acquired using a point monitor placed at the hot spot regions imply that the CTP mode has longer lifetime and it oscillates 180 ° out of phase with both LSP and BDP modes (see **Figure 2(c, d)**).

To quantitatively support this claim, we obtained the amplitude of the electric near-field decay by fitting the simulation data to the exponential decay function equation of $E(t) = E_0 e^{-t/\tau}$ [24]. Accordingly, the lifetime of CTP mode is found to be longer ($\tau_{CTP} = 13.80\ fs$) compared to those of LSP and BDP modes ($\tau_{LSP} = 10.36\ fs, \tau_{BDP} = 7.20\ fs$) (see **Figure 2(e)**). These values, which are consistent with the dephasing times obtained indirectly from the scattering spectra, imply that CTP mode has a longer lifetime and thus may find applications in designing ultrasensitive sensors.

Finally, using the inverse relationship of the CTP resonance frequency and charge transfer rate, we determined the charge transfer time $t_{CT} \sim \pi/2\omega_{CTP} \sim 1.02\ fs$. This time scale is about five times faster than the charge transfer rate between two 40 nm Au nanoparticles with a 38 nm gap mediated by Ag nanoparticle that bridges the gap and transfers energy coherently between the two Au nanoparticles [27]. As a result, the energy dissipation in the conductive junction is small, which shows the low-loss mechanism of information and ultrafast transfer of charges between conductively coupled particles.

*Nanojunction volume effects:* To manipulate the electron transport across conductive junction, first, we modulated the nanolinker volume by modifying its thickness as well as length and simulated the scattering spectra of conductively linked nanodisk dimer as a function of junction volume (see **Figure 3(a)**). As it can be easily deduced from the theoretical linear relationship between junction conductance and its cross section shown in equation (1), increasing the nanojunction width for a fixed length (*l* = 30 nm) induces enhanced charge flow across the junction and results in significant shift of the CTP peak towards higher energies [6]. On the other hand, increasing length of the nanojunction by maintaining its width constant (*w* = 22 nm), the CTP resonance energy gradually redshifts to lower energies because of longer time it takes for charges to transport across longer junctions [6, 9]. To confirm this delayed charge transport time effect, using equation (3), we have also calculated the charge transfer rate for each junction length and it is found to be within the range of 0.95 fs to 1.05 fs. Nevertheless, for the studied junction volume effects, the CTP resonance shift for the thickness effect is prominent ($\hbar\Delta\omega_{res} \approx 206\ meV$) compared to that of the length effect which is about 116 meV. This can be attributed to the enhanced junction mode volume due to the thickness effect that increases its volume in 2D unlike the 1D effect of the junction length.

Furthermore, the effect of junction mode volume on the CTP spectral linewidth is obtained by fitting scattering spectra using a Lorentzian function. The CTP resonance linewidth shows broadening as the nanojunction width increases (**Figure 3(b)**). Consequently, the dephasing time of CTP mode obtained from the scattering spectra fitting reveals a monotonic decay at ultrafast time scale, decreasing from 12.92 fs to 8.32 fs with increasing the junction thickness from 22 nm to 34 nm (see **Figure 3(c)**). This can be attributed to the increased radiative damping associated to increase in the junction volume. Because, in principle, the radiative damping rate of particle plasmon is directly proportional to the particle volume [*17*]. In contrast, the increment of nanojunction volume via its length has reverse effect both on the CTP linewidth and dephasing time. For the studied junction length range (*l* = 20 nm to *l* = 35 nm), the CTP linewidth decreases as the junction volume increases. Conversely, the CTP dephasing time increases from 11.74 fs to 13.85 fs as the junction length increases (see **Figure 3(d)**). This peculiar effect of nanojunction length can be attributed to the counter-intuitive relationship between near-field enhancement and dephasing time. Generally, plasmonic dimers with small gap sizes show stronger near-field enhancement and shorter dephasing times owing to enhanced radiative decay [23].

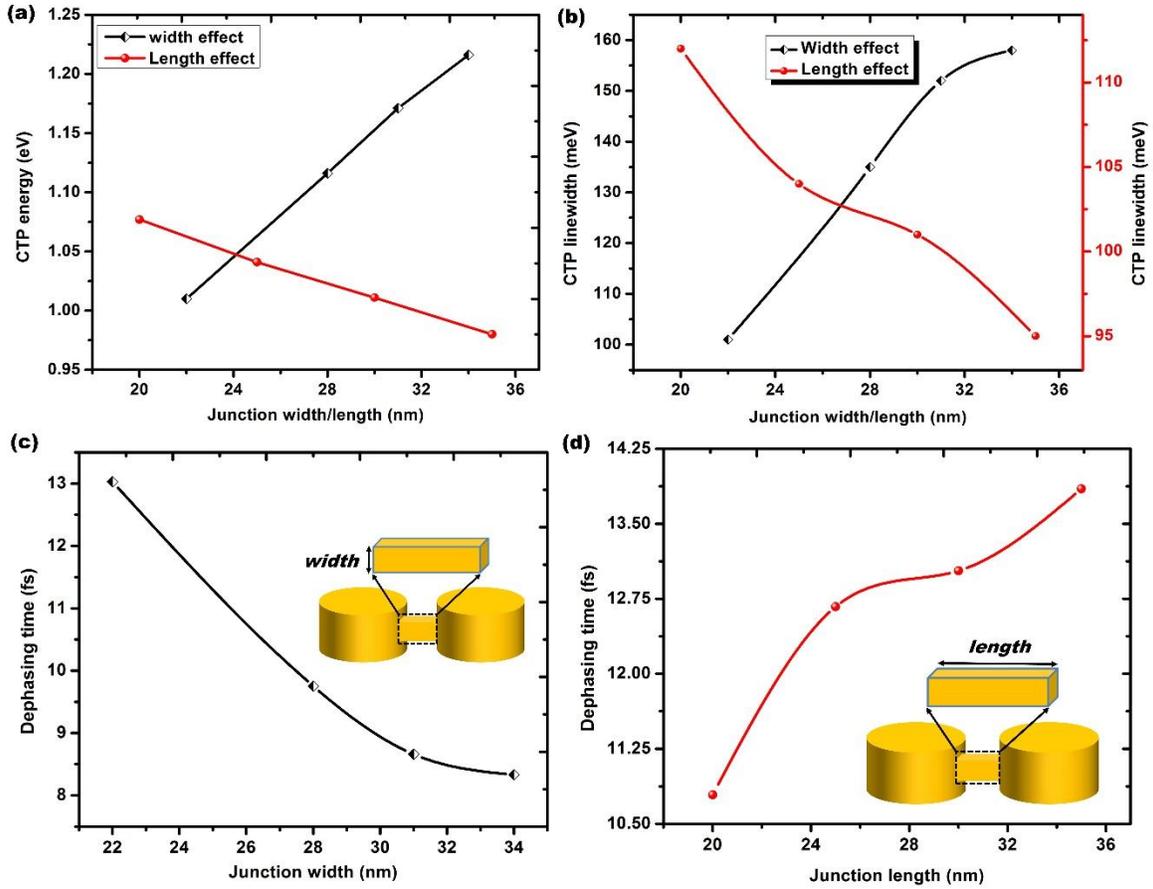

**Figure 3**. The effect of junction mode volume on the CTP resonance energy, spectral linewidth and dephasing time. (a) CTP energy shift due to increase in the junction width and length. (b) CTP spectral linewidth signatures as function of junction width/length. (c, d) Corresponding CTP dephasing times as a function of junction width and length respectively.

To better understand the peculiar effect of the nanojunction mode volume on the CTP linewidth and dephasing time, we have obtained near-field spatial maps and temporal profiles of conductively linked nanodisk dimers with different junction lengths and widths (see **Figure 4**). The near-field maps shown in **Figure 4(a)** suggest that for thicker junction widths ($w = 34$ nm), which are comparable to the nanoparticle height (H = 35 nm), the near-field distribution of CTP mode looks like LSP mode of single nanorod. Similarly, for longer junction lengths (for example, $l = 35$ nm), the near-field enhancement in the gap region tends to decline (see **Figure 4(b)**). The reduction in the near-field enhancement due to increase in junction length may be attributed to weaker coupling of the nanoparticles and reduced near-field localization [23].

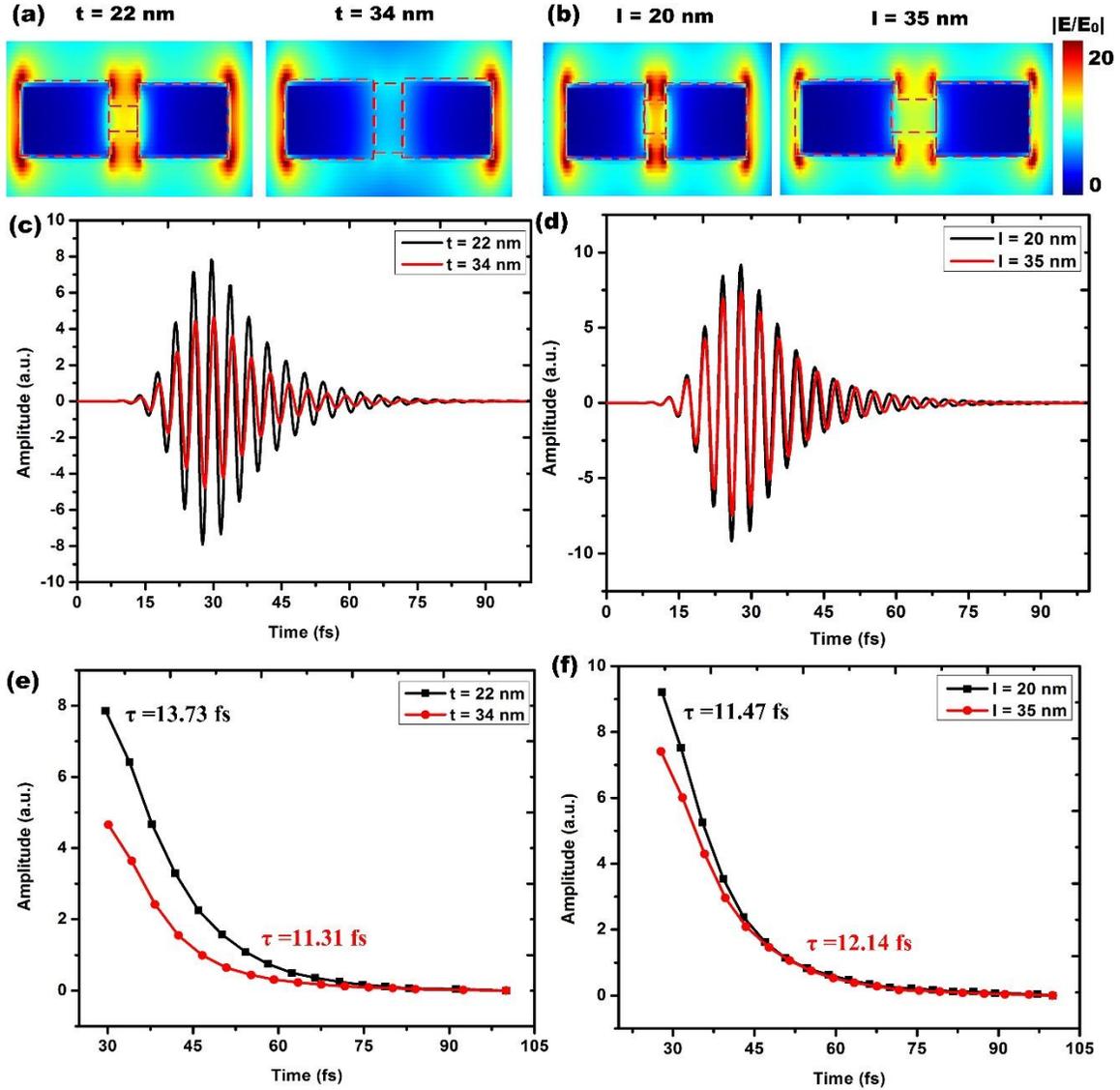

**Figure 4**. (a, b) Spatial maps of resonantly excited CTP modes simulated for various junction thicknesses and lengths. (c, d) Corresponding electric near-field amplitudes obtained at 1 nm away from the nanojunction. (e, f) CTP induced near-field decay rates as functions of junction thickness and length.

In the same fashion, the near-field amplitude of CTP modes calculated at 1 nm away from the junction imply that increasing the junction volume induces decrease in its near-field amplitude, which can be attributed to weaker coupling of the nanoparticles (**Figure 4(c, d)**). However, the near-field decay rates obtained from exponential fitting show distinct behaviors. While the near-field lifetime declines from 13.73 fs to 11.31 fs as the junction thickness increases, conversely, it tends to increase from 11.47 fs to 12.14 fs as the junction length increases (see **Figure 4(e, f)**). In fact, it has been shown that gap plasmon near-field lifetime gets longer as the inter-particle separation gets larger, reaching a maximum value at a certain gap and then starts to decline to the level of isolated nanoparticles [*24*]. These findings signify the effect of nanojunction mode volume in manipulating charge transfer in linked plasmonic nanoparticles and thus might advance recent efforts to understand how junction nature impacts the conduction mechanism in nano-networks [29].

*Nanoparticle size effects:* As implied in equation (2), the size of nanoparticles in conductively linked nanosystems directly affect CTP resonance. Thus, we explore the impact of nanodisk size (diameter) on the linewidth and dephasing time of CTP mode. From **Figure 5 (a, b),** one can note that both spectral linewidth and dephasing time are drastically alterted when the nanoparticle diameter is altered from D = 37 nm to D = 97 nm. For the simulated nanodisk diameters, the CTP linewidth decreases monotonously from 184.37 meV to 101.82 meV ($\Delta\Gamma \approx 82.55\ meV$) (see **Figure 5(a)**). This is due to the fact that increasing the nanoparticle size increases charge separation distance, resulting in longer charge transport time [6, 9].

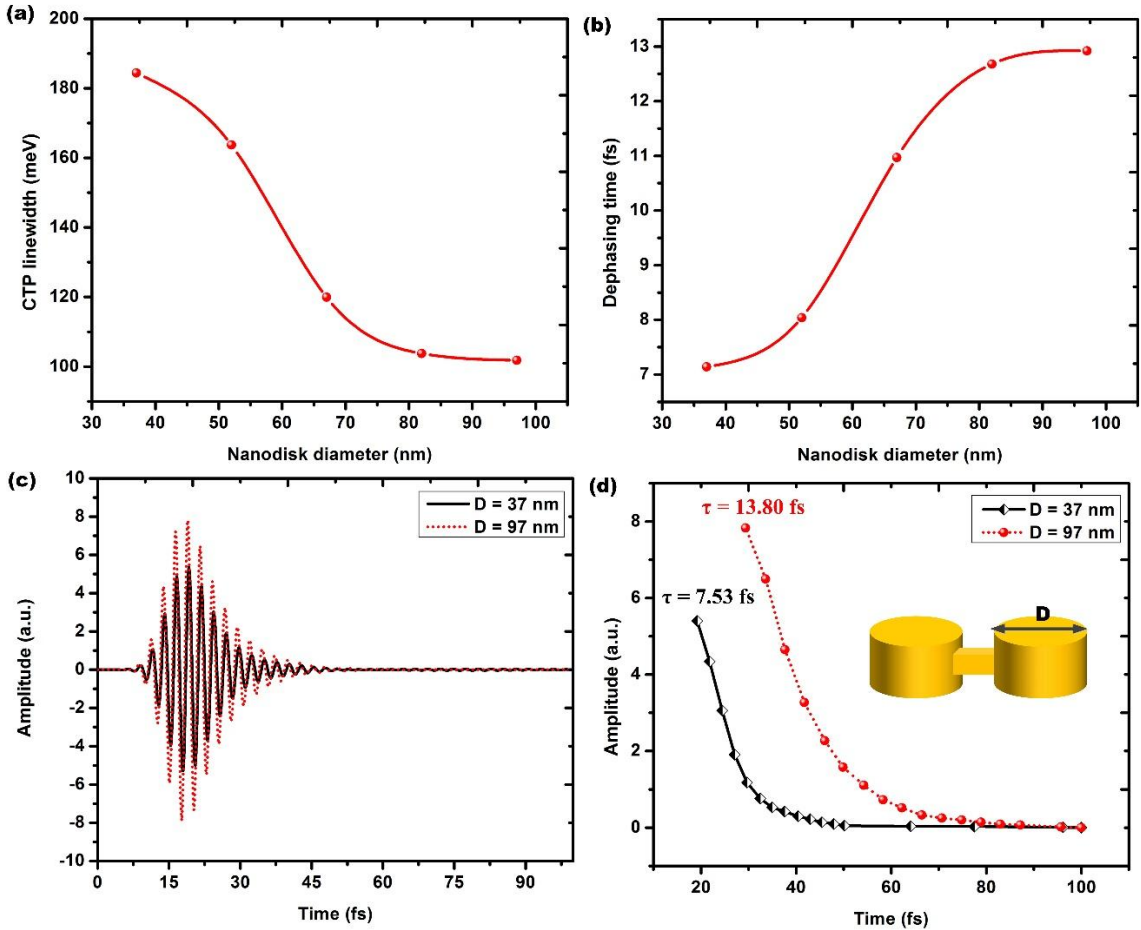

**Figure 5**. (a, b) The effects of nanodisk diameter on CTP linewidth and dephasing time. (c, d) Near-field amplitudes and decay rates calculated for various nanodisk diameters.

This fact is confirmed by calculating the charge transport time for the studied nanodiask diameters, which increases from 0.65 fs to 1.02 fs as the nanodisk diameter increases from 37 nm to 97 nm. In the same manner, the dephasing time of the CTP resonances acquired from the scattering spectra fitting shows notable increment, rising from 7.14 fs to 12.92 fs (see **Figure 5(b)**). It is worth noting that the dephasing time of the linked nanodisk dimer for D = 37 nm is comparable with the dephasing time of the nanorod obtained from the scattering spectrum shown in Figure 1(b). In fact, for the linked nanodisk dimer where the disk diameter is comparable to its height (D = 37 nm, H = 35 nm), the connected nanodimer shows a single resonance peak and its local field distribution resembles that of a nanorod. As a result, the CTP mode behaves like LSP mode and thus it decays at a faster rate for smaller nanodisk diameters.

To further support the dephasing time results acquired from far-field spectra, we have also obtained the near-field decay rates calculated at 1 nm away from the nanojunction for two extreme nanodisk sizes of D = 37 nm and D = 97 nm(see **Figure 5(c, d)**). The near-field amplitude decays at a faster rate for smaller nanodisk dimers than larger ones, extending from 7.53 fs to 13.80 fs as the nanodisk size increases, which is consistent with the CTP dephasing times acquired from the scattering spectra. These results imply that, to sustain a desired CTP mode in conductively coupled nanoparticles, both the junction conductance and nanoparticle sizes should be carefully designated.

4. Conclusions

As shown in this study, the dephasing time of charge transfer plasmon resonances obtained from scattering spectra fitting is found to be about 7 fs to 13 fs, which is further verified by calculating the near-field decay rates. Our findings are within the theoretical limit of plasmon dephasing time that occurs on the order of 5−20 fs [19]. For the thickest nanojunction width that is comparable to the nanodisk height, the connected nanosystem resembles a nanorod and its CTP resonance behaves like particle plasmons. However, for the plasmonic nanodimers connected by thinner conductive nanojunctions, CTP mode has an intense resonance that redshifts as the junction length increases. We find that the CTP lifetime increases monotonously as the junction length increases, which is a typical behavior of bonding dimer plasmons [23, 24]. This implies that CTP modes behave as dimer plasmons when the conductive linker is thinner and longer.

**Table 1**. Spectral and temporal signatures of CTP modes in conductively connected Au nanodisk dimers at various junction geometries.

| Junction geometry | $\hbar\omega_{CTP}$ (eV) | $\Gamma$ (meV) | $T_2$ (fs) | $\tau_{CTP}$ (fs) | $t_{CT}$ (fs) |
|---|---|---|---|---|---|
| $l = 35$ nm, $t = 22$ nm | 0.980 | 95.10 | 13.85 | 12.71 | 1.05 |
| $l = 30$ nm, $t = 22$ nm | 1.011 | 101.82 | 12.92 | 12.57 | 1.02 |
| $l = 20$ nm, $t = 22$ nm | 1.077 | 112.04 | 11.74 | 11.82 | 0.95 |
| $t = 34$ nm, $l = 30$ nm | 1.216 | 158.09 | 8.32 | 11.24 | 0.85 |

Regardless of these similarities, compared to the well-known particle plasmons and dimer plasmons, we find that CTP modes have some peculiar characteristics. First, for equivalent mode volumes of studied nanostructures, CTP modes have a narrow linewidth ($\Delta \Gamma \approx 48\ meV$). As a result, CTP modes have a superior Q factors ($Q_{CTP} \approx 10$, $Q_{LSP} \approx 6.8$) implying that CTP modes are ideal candidates for developing ultrasensitive sensors. Furthermore, the dephasing time of CTP mode is about twice as large as the bonding dimer mode and a few femtoseconds longer (about 3 fs) than that of LSP mode. Another important observation is that the electric near-field amplitude of CTP mode shows about 180° out-of-phase oscillation with both the LSP and BDP modes. Furthermore, the typical charge transfer rate calculated for conductively coupled Au nanoparticles is found to be about 1 fs. Overall, the calculated CTP lifetime and charge transfer rate are both on the femtosecond timescale (see **Table 1**), enabling efficient and coherent transfer of energy between conductively connected nanoparticles [27].

In summary, we have explored the spectral signatures and temporal dynamics of charge transfer plasmons in conductively connected gold nanodisk dimers. Using classical electromagnetic simulations, we obtained the scattering spectra of linked nanodimers with intense CTP resonances in the near infrared regime. We used the Lorentzian function to fit the scattering spectra of CTP resonances and investigated their spectral linewidths for different nanostructure parameters. Moreover, using the Heisenberg's energy-time relationship, we quantitatively determined the dephasing times of CTP modes, which are further confirmed by calculating the optical near-field decay rates acquired at 1 nm away from the conductive junction. Our results based on the far-field spectral fittings and near-field analysis imply that, compared to the particle plasmons and dimer plasmons, charge transfer plasmons have extended lifetime. We have also shown that the spectral linewidth and dephasing time of CTP modes in conductively coupled plasmonic nanoparticle dimers can be further extended by controlling the geometric parameters of nanolinker and nanoparticles. These findings imply that conductively connected nanoparticles hold great promise as channels for coherent transfer of energy and information and thus we anticipate that our findings might have implications for building molecular sensors and ultrafast optoelectronic devices with interconnected plasmonic nanostructures.